\documentclass[usenatbib]{mnras}
\usepackage{epsfig}
\usepackage{amsmath,amssymb}
\usepackage{color}
\usepackage{booktabs}
\usepackage{pdflscape}
\usepackage[strict]{changepage}

\usepackage{epsfig}
\usepackage{color}
\usepackage{aas_macros}  
\usepackage{amsmath,amssymb}
\usepackage{chngcntr}
\usepackage[normalem]{ulem}  
\graphicspath{{figures/}}

\newcommand{\Msolar}{\mbox{\,$\rm M_{\odot}$}}        

  \newcommand{\Teff}{\mbox{\,\em T$_{\rm eff}$}}         
  \newcommand{\sg}{\mbox{\,log $g$}}                     
 \newcommand{\teff}{\mbox{\,$T_{\rm eff}$}}      
\newcommand{\lgcs}{\mbox{\,$\log g / {\rm cm\,s^{-2}}$}}        

  \newcommand{\nHe}{\mbox{\,$n_{\rm He}$}}               
  \newcommand{\vsini}{\mbox{\,$v\,\sin i$}}              
  \newcommand{\kmsec}{\,\mbox{$\mbox{km}\,\mbox{s}^{-1}$}}    
  \def\simge{\mathrel{\raise1.16pt\hbox{$>$}\kern-7.0pt
    \lower3.06pt\hbox{{$\scriptstyle \sim$}}}}           
  \def\simle{\mathrel{\raise1.16pt\hbox{$<$}\kern-7.0pt
    \lower3.06pt\hbox{{$\scriptstyle \sim$}}}}           
\newcommand{\ledz}{UVO\,0825+15}   
\newcommand{\crimson}{LS\,IV$-14^{\circ}116$}  

\title[Heavy-metal subdwarfs]{A search for heavy-metal stars: abundance analyses of hot subdwarfs with Subaru\thanks{based on observations made with the Subaru Telescope}
}
\author[Naslim N. et al.]{Naslim N.$^{1,2,3}$\thanks{E-mail: naslim.n@uaeu.ac.ae},
C. S. Jeffery$^{3,4}$\thanks{E-mail: Simon.Jeffery@armagh.ac.uk} and V. M. Woolf$^{5}$ \\
$^{1}$Department of Physics, United Arab Emirates University, Al-Ain, UAE, 15551\\
$^{2}$Academia Sinica Institute of Astronomy and Astrophysics, Taipei 10617, Taiwan \\
$^{3}$Armagh Observatory and Planetarium, College Hill, Armagh BT61 9DG, UK\\
$^{4}$School of Physics, Trinity College Dublin, College Green, Dublin 2, Ireland\\
$^{5}$Physics Department, University of Nebraska at Omaha, 6001 Dodge St, Omaha, NE, 68182, USA\\
}

\date{Accepted \ldots. Received \ldots; in original form \ldots}

\begin{document}

\pagerange{\pageref{firstpage}--\pageref{lastpage}} \pubyear{2019}

\maketitle

\label{firstpage}

\begin{abstract}
The discovery of extremely zirconium- and lead-rich surfaces amongst a
small subgroup of hot subdwarfs has provoked questions pertaining to
chemical peculiarity in hot star atmospheres and about their
evolutionary origin.  With only three known in 2014, a limited search
for additional `heavy-metal' subdwarfs was initiated with the Subaru
telescope. Five hot subdwarfs having intermediate to high surface
enrichment of helium were observed at high-resolution and analyzed for
surface properties and abundances. This paper reports the
  analyses of four of these stars.
PG\,1559+048 and FBS\,1749+373, having only
intermediate helium enrichment, show strong lines of triply ionized
lead.  PG\,1559+048 also shows a strong overabundance of germanium and
yttrium. With more helium-rich  surfaces,
Ton\,414 and J17554+5012, do not show evidence of heavy-metal
enrichment.  This limited survey suggests that extreme enrichment of
`heavy metals' by selective radiative levitation in hot subdwarf
atmospheres is suppressed if the star is too helium-rich.  
\end{abstract}


\section{Introduction}
\label{s:intro}

The origin of hot subluminous stars which have surfaces rich in
heavy metals including zirconium and lead poses a challenge.  They
have surfaces dominated neither by hydrogen
nor by helium and belong to
a group known as intermediate helium subdwarfs (iHe-sds) \citep{naslim10}. By
convention, this group is characterized by a surface helium
abundance (fractional abundance by number) $0.1 < n_{\rm He} <
0.9$\footnote{$\equiv -0.95 < \log y < +0.95$, where $y\equiv n_{\rm
    He}/n_{\rm H}$} and corresponds approximately to a
\citet{drilling13} helium class in the range He15 -- He35
(ibid. Fig.\,12). 
Surface temperatures and gravities cover  the ranges
  $30\,000 < T_{\rm eff}/{\rm K} < 42\,000$ and $5.0 <
  \log g/ (\rm cm\,s^{-2}) < 6.0$. 

In contrast, normal subdwarf B (sdB) stars have helium-poor atmospheres
  ($n_{\rm He}<0.1$) whilst helium-rich hot subdwarfs (He-sds)
  show surface helium abundance in a range $0.9 < n_{\rm He} < 0.99$
  with a wide range of surface temperature $T_{\rm eff}=25\,000-45\,000$\,K \citep{naslim10}.
  
During a photometric survey of 21 iHe-sds \citep{ahmad04a}, \citet{ahmad05b} discovered evidence for g-mode pulsations in \crimson, later confirmed by \citet{green11} and \citet{jeffery11.ibvs}. 
Follow-up spectroscopy of \crimson\ led to the discovery of 3--4\,dex overabundances of the trans-iron elements zirconium, yttrium, strontium and germanium \citep{naslim11}, which  had not previously been found in other iHe-sds.  
A targeted survey of iHe-sds subsequently led to the discovery of two stars, HE\,2359--2844 and HE\,1256--2738, with $\approx 4$ dex overabundances of lead, together with other heavy elements \citep{naslim13}.  
More recent discoveries include Feige\,46 -- a pulsating `twin' to \crimson\,\citep{latour19a}, and EC\,22536--4304, a new lead-rich subdwarf \citep{jeffery19b}.  
The over-abundance of trans-iron elements has led to these unusual stars becoming known as 'heavy-metal subdwarfs'.

The questions raised by these discoveries include the numerical
frequency of heavy-metal subdwarfs amongst the population of hot
subdwarfs as a whole, the range of effective temperature and surface
gravity over which they occur, whether there are systematics within the group 
regarding which elements are enhanced, whether they occur in binaries, the
evolution channel which produces subdwarfs with 'intermediate'
helium-rich surfaces, and the mechanism by which the surfaces become
metal-rich. One hypothesis is that iHe-sds are in transition phase from helium-rich
  to helium-poor due to atmospheric diffusion processes. Atmospheric
  helium gradually sinks within a time-scale of 10$^5$ years as the
  helium-rich star contracts towards the zero-age horizontal
  branch. In their radiation-dominated atmosphere, certain selected
  species concentrate into specific regions where opacities are high
  enough to observe extreme over-abundances. A clear theoretical account
  of this process is still lacking.
  
This paper starts to address the first two questions by reporting the
results of a limited spectroscopic survey of helium-rich subdwarfs
made using the High Dispersion Spectrograph (HDS) of the Subaru
Telescope.  One star observed in the survey, \ledz\, has already been
reported as a lead-rich subdwarf \citep{jeffery17a}.  In \S\,2, we
report the remaining observations. In \S\,3 and 4 we report the measurements of atmospheric parameters and abundance analyses, and in \S\,5 we draw interim conclusions.

\begin{figure*}
\begin{center}
\includegraphics[width=.99\textwidth,angle=0]{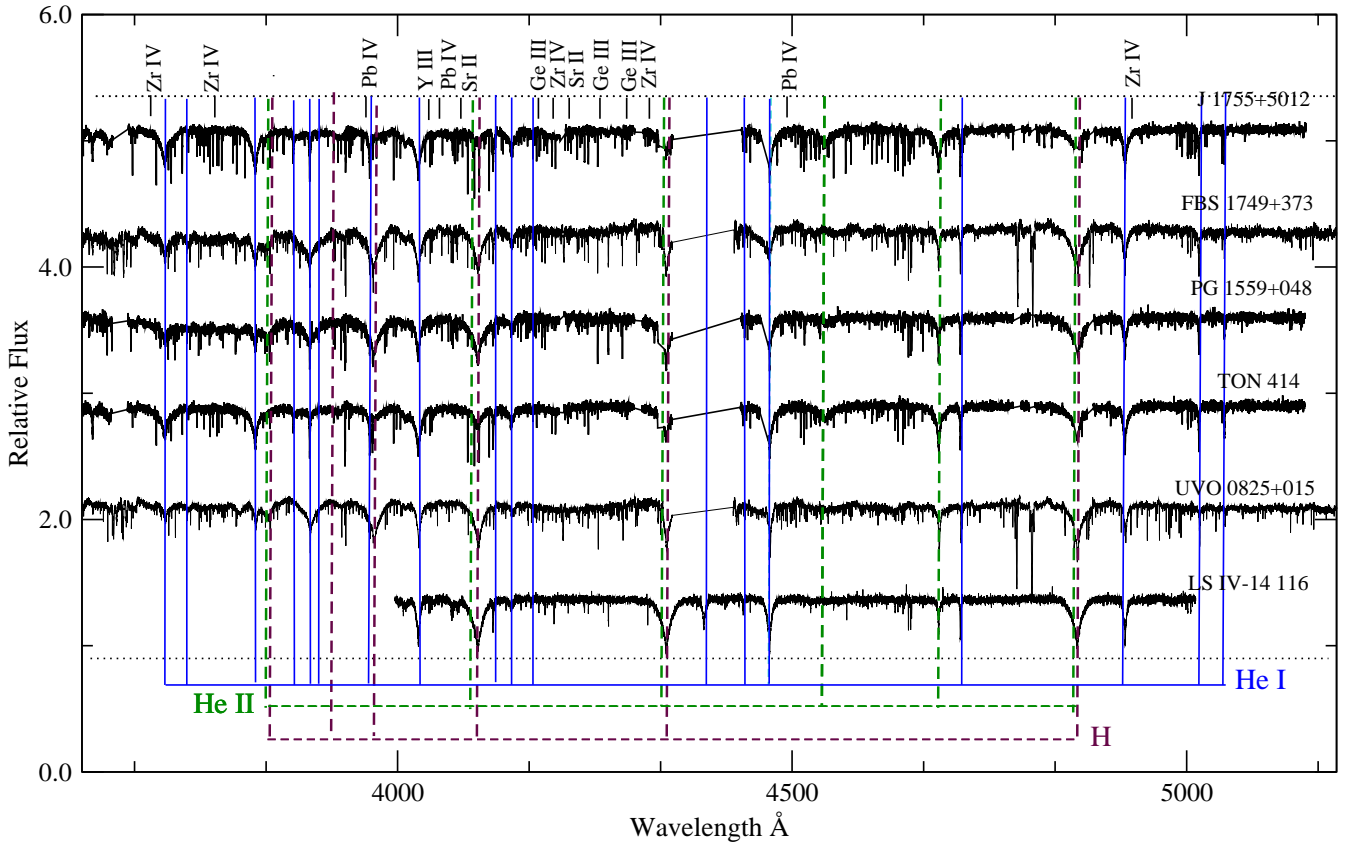}
\end{center}
\caption{Subaru/HDS spectra of five helium-rich subdwarfs together with (bottom) the AAT/UCLES spectrum of \crimson \,from \citet{naslim11}. Locations of pricipal H, He{\sc i}, He{\sc ii}, Ge{\sc iii}, Y{\sc iii}, Zr{\sc iv} and Pb{\sc iv} lines are indicated.}
\label{fig:hds}
\end{figure*}

\section[]{Observations}
\label{s:obs}

Observations of five He-sds were obtained on 2015 June 3 with the HDS
on the Subaru telescope in service mode (program S15A-206S). 
The sample analysed in this paper comprised PG\,1559+048, Ton\,414, FBS\,1749+373, and
GALEX\,J175548.50+501210.77.   
These stars were identified as He-sds sufficiently bright for high-resolution high
signal-to-noise spectroscopy, and with $T_{\rm eff}$ in a range
similar to that in which heavy-metals had been discovered previously
\citep{naslim11,naslim13}.

Two exposures were made consecutively for each star, with exposure times
of 1200\,s in each case, except for the brighter GALEX
J175548.50+501210.77, where an exposure times of 300\,s
was used. A slit width of 0.4 mm was used,
corresponding to a projected resolution $R = 45\,000$. 
The data were reduced as described by \citet[][\S\,2.3]{jeffery17a}. 
The Subaru HDS spectra of the current sample are shown in Figure\,\ref{fig:hds}, 
together with the fifth sample member UVO\,0825+15 \citep{jeffery17a} and the AAT spectrum of LS\,IV$-14^{\circ}116$ \citep{naslim11}.

Radial velocities ($v_{\rm r}$) measured for all four stars were reported by \citet{martin17a}. However those measurements omitted a correction to the heliocentric rest frame, and should have been given as in Table 1.

PG\,1559+048 (= GALEX\,J160131.30+044027.00) is classified sdOA by
\citet{green86}, sdBC0.2VII:He22 by \citet{drilling13}, and He-sdB by
\citet{nemeth12}, thus indicating an intermediate helium abundance.
Our Subaru/HDS spectrum shows relatively strong N\,{\sc ii}, N\,{\sc
  iii}, C\,{\sc ii}, and C\,{\sc iii} lines, weak Si\,{\sc iii},
Si\,{\sc iv}, S\,{\sc iii}, and Ca\,{\sc iii} lines, together with
strong hydrogen Balmer, He\,{\sc i} and He\,{\sc ii} 4686\,\AA\ lines.
Also identified are: Ti\,{\sc iii} 3915.26\,\AA, Ti\,{\sc iv} 4618.04,
4677.59\,\AA, V\,{\sc iv} 4841.26, 4985.64\,\AA, Y\,{\sc iii}
4039.602, 4040.112\,\AA, Pb\,{\sc iv} 4049.80, 4496.15 and
3962.48\,\AA.  The lead lines are the same as those identified in
HE\,2359--2844 and HE\,1256--2738 \citep{naslim13}, whilst the yttrium
lines are the same as those found in LS\,IV$-14^{\circ}116$
\citep{naslim11} and HE\,2359--2844.

FBS\,1749+373\,(=\,GALEX\,J175137.4+371952)\,was identified as a blue
stellar object in the First Byurakan spectral sky Survey (FBS)
\citep{abrahamian90.II} and updated to sdOB by \citet{mickaelian08}
and to He-sdB by \citet{nemeth12}. The latter noted that the amplitude of $v_{\rm r}$ relative to the local standard of rest $|v_{\rm r}({\rm LSR})|>100\kmsec$ and also that the star's parameters lie in the domain of rapid sdB pulsators. Our
Subaru/HDS spectrum shows  H, He, C, N, Si and Ca lines similar to
PG\,1559+048.  It also shows Pb\,{\sc iv} 4049.80\,\AA, but Pb\,{\sc
  iv} 3962.48 and 4496.15\,\AA\ are not detected. Cl\,{\sc ii}
3720.45\,\AA, weak Ti\,{\sc iii} 3915.26\,\AA\ and Ti\,{\sc iv}
4677.59\,\AA\ are also seen. The heliocentric velocity of $-69.7\kmsec$ appears different to the lower limit indicated by \citet{nemeth12}. However, the latter did not state whether $v_{\rm r}({\rm LSR}) < -100$ or $>+100\kmsec$ and their spectral resolution $\delta \lambda > 5$\AA points to a likely minimum velocity error $\delta v = (c\delta \lambda/\lambda) / 10 > \pm 35\kmsec$ (assuming $1/10$ pixel precision). Whilst membership of a spectroscopic binary cannot be completely excluded, there is no evidence for a 3$\sigma$ difference between the two existing velocity reports. 

Ton\,414 (=\,PG\,0921+311\,=\,GALEX\,J092440.1 +305013) was
  classified sdOB by \citet{green86}, and He-sdO by
  \citet{thejll94}. The Subaru HDS spectrum of Ton\,414 shows weak C\,{\sc
    iii} 4647 and 4650\,\AA\, lines, Si\,{\sc iii}, Si\,{\sc iv},
  S\,{\sc ii} and S\,{\sc iii} lines. This star shows relatively
  strong N\,{\sc ii}, N\,{\sc iii}, He\,{\sc i} and He\,{\sc ii} 4686\,\AA\ lines.
  
  GALEX\,J175548.50+501210.77\,(=\,TYC\,3519-907-1\,=\,J$17554+5012$
  hereafter) was classified He-sdB by \citet{nemeth12}. The Subaru HDS
  spectrum shows strong He\,{\sc i}, He\,{\sc ii} 4686\,\AA\ lines,
  N\,{\sc ii}, N\,{\sc iii}, S\,{\sc ii} and S\,{\sc iii} lines along
  with relatively weak C\,{\sc iii}, Si\,{\sc iii}, and Si\,{\sc iv}
  lines. Neither Ton\,414 nor J17554+5012 show any detectable
  heavy-metal absorption lines. 
  Ne\,{\sc ii} and Al\,{\sc iii} lines are seen in both stars.

\begin{table*}
\caption{Atmospheric parameters for the programme stars.}
\label{t_pars}
\begin{tabular}{@{}lllllllll}
\hline
Star & Grid & $T_{\rm eff}/(\rm K)$ & $\log g/{\rm cm\,s^{-2}}$ & $n_{\rm He}$  & $\log y$ & $v \sin i$ & $v_{\rm r}$ & Source\\
     &   &                      &          &            &      & $({\rm km\,s^{-1}})$ & $({\rm km\,s^{-1}})$ \\
\hline
PG\,1559+048 &  m10 & $37\,200\pm1600$   & $6.00\pm0.15$ & $0.20\pm0.03$ & $-0.60\pm0.08$ &   $2\pm0.3$    & $-34.0\pm0.9$ & 1     \\
PG\,1559+048 &  sdb &$37\,120\pm1600$   & $5.98\pm0.15$ & $0.22\pm0.02$ & $-0.55\pm0.05$ &        &     & 1     \\
   &  & $40\,330\pm860$         & $6.16\pm0.18$        &    &  $-0.53\pm0.21$   &  &  & 2 \\[1mm]

FBS\,1749+373&   m10 &$36\,800\pm2000$   & $5.80\pm0.20$ & $0.27\pm0.06$ &  $-0.43\pm0.14$  &  $5\pm2$ & $-69.7\pm 0.2$       & 1    \\
FBS\,1749+373&   sdb &$36\,500\pm1800$   & $5.85\pm0.22$ & $0.28\pm0.05$ &  $-0.41\pm0.11$  &      &     & 1    \\

         &   & $34\,630\pm600$      &  $5.89\pm0.12$    &$0.34\pm0.03$ &$-0.28\pm0.06$ & &  & 2 \\[1mm]

Ton\,414&   m10 &$37\,200\pm1000$   & $5.65\pm0.3$ & $0.79\pm0.10$  & $+0.58\pm0.24$ & $2\pm1$  & $-22.7\pm0.5$      & 1    \\
     &     & $40\,830$ & $5.84\pm0.15$&    &    &    &  & 2 \\
     &     & $41\,000$ & $5.4$  &  $0.85$  & $+0.75$  &   &   & 3\\[1mm]

J17554+5012&   m10 &$39\,500\pm1500$   & $5.70\pm0.10$ & $0.95\pm0.05$   & $+1.27\pm0.35$ &   $3\pm2$  & $-60.3\pm0.2$       & 1    \\
    &      & $40\,370\pm940$         & $5.96\pm0.15$  &  &  $+1.30\pm0.27$  &  &     &    2\\

\hline
\end{tabular}\\
\parbox{170mm}{
Reference: 
1. This paper. $v_{\rm r}$ are as given by \citet{martin17a} corrected to the heliocentric rest frame.  
2. \citet{nemeth12}
3. \citet{thejll94}
}
\end{table*}

%
%

\begin{table*}
\centering
\caption{Elemental abundances in the form  $\log \epsilon_i$ with errors in parentheses and upper limits indicated by ``$<$''.}
\label{t:abs}
\setlength{\tabcolsep}{1.7pt}
\begin{tabular}{@{\extracolsep{0pt}}p{23mm}l lll lll lll ll}
\hline
Star 					& H			& He 		& C			& N			& O			& Ne		& Mg 		& Al		& Si		& S 			\\
\hline 
PG\,1559+04\,$^a$   & 11.89  & 11.30  & 8.65(0.47)  &  8.08(0.23) &  $<7.12$  &  $<8.0$  &  $<7.10$  &  6.64(0.22)  & 6.57(0.69)  &  7.84(0.40)   \\
PG\,1559+04\,$^b$     &  11.89     &   11.30     & 8.64(0.46)  &  8.06(0.22) & $<7.12$    & $<8.0$  &   $<7.10$ &   6.62(0.22) &   6.54(0.66) &   7.80(0.39)     \\
FBS\,1749+373\,$^c$ & 11.76  &  11.33 &  8.62(0.32) & 8.39(0.35)  & $<7.10$   & $<8.60$  & $<7.00$   & $<5.80$   &  6.69(0.40) & 7.40(0.29)    \\ 
FBS\,1749+373\,$^d$ &   11.76    &   11.33     & 8.63(0.32)  &  8.40(0.35) & $<7.10$    & $<8.40$  &  $<7.00$    &  $<5.80$  &  6.70(0.39) & 7.41(0.29)               \\
Ton\,414\,$^e$  & 11.00   & 11.60  & 6.94(0.16)  &  8.42(0.26) &  $<7.30$  & 8.28(0.48)  &  $<6.50$  &  6.27(0.30)  & 6.95(0.36)  &  6.72(0.30)     \\
J17554+5012\,$^f$   & 10.40  & 11.58  & 6.94(0.37)  &  8.78(0.30) &  7.99(0.19)  & 8.50(0.67)  & $<7.00$   &  6.55(0.37)  & 7.55(0.30)  &  7.19(0.39) &    \\[1mm]
Sun$^{1}$				& 12.00		& 10.93		& 8.43		& 7.83		& 8.69		& 7.93 		&  7.60		& 6.45		& 7.51		& 7.12					    \\[3mm]
\hline
Star			& Cl  	 	&	Ar	& Ca 		& Ti 		& V 				& Fe						& Ge		& Y	& Zr	& Pb \\
\hline
PG\,1559+048$^a$   &  6.63(0.24)& $<8.80$&  7.87(0.24)  & 7.86(0.42)  & 7.40(0.32)    & $<7.50$   &       6.69(0.30)   & 6.09(0.16)& $<6.24$ & 5.10(0.14) \\
 PG\,1559+048$^b$ &   6.59(0.24) & $<8.70$&  7.88(0.23)  & 7.86(0.41)  & 7.43(0.32)    & $<7.50$   &       6.62(0.30)   & 6.07(0.16)& $<6.21$ & 5.08(0.14)      \\     
FBS\,1749+373$^c$  & 6.14(0.24) & $<8.00$ &  7.83(0.32)  & 7.98(0.41)  & $<7.15$     & $<7.20$   &   $<5.32$       & $<5.30$ & $<6.03$& 4.89(0.16) \\ 
FBS\,1749+373$^d$  & 6.14(0.24)  & $<8.00$  &  7.84(0.32)    & 7.98(0.40)  & $<7.15$    &  $<7.20$ & $<5.32$   &      $<5.30$   & $<6.03$   &  4.89(0.16)           \\
Ton\,414$^e$ & $<5.80$  & $<7.75$ &  $<7.50$  &  $<7.30$   & $<6.90$  &     $<7.15$  &   $<5.40$      & $<5.19$& $<5.93$ & $<4.25$ \\
J17554+5012$^f$  & $<5.75$  & $<7.80$ &  $<7.40$  &  $<7.15$   & $<6.75$    & $<7.05$   &  $<5.40$       & $<5.13$& $<5.88$ & $<4.18$ \\[1mm]
Sun$^{1}$  		& 5.50	&	6.40	& 6.34 		& 4.95 		& 3.93 				& 7.50       		& 3.65		& 2.04	& 2.58	& 1.75 \\
\hline
\end{tabular}\\
\parbox{170mm}{
Notes:\\
$a$:   model: $T_{\rm eff} =  38\,000$\,K, $\log\,g=6.0$, $n_{\rm He}=0.200$, {\bf m10}                    \\
$b$:   model: $T_{\rm eff} =  38\,000$\,K, $\log\,g=6.0$, $n_{\rm He}=0.200$, {\bf sdb} \\
$c$:   model: $T_{\rm eff} =  36\,000$\,K, $\log\,g=6.0$, $n_{\rm He}=0.300$, {\bf m10}                    \\
$d$:   model: $T_{\rm eff} =  36\,000$\,K, $\log\,g=6.0$, $n_{\rm He}=0.300$, {\bf sdb} \\
$e$:   model: $T_{\rm eff} =  38\,000$\,K, $\log\,g=6.0$, $n_{\rm He}=0.699$, {\bf m10}                    \\
$f$:   model: $T_{\rm eff} =  40\,000$\,K, $\log\,g=6.0$, $n_{\rm He}=0.949$, {\bf m10}                    \\
1. \citet{asplund09}; photospheric except helium (helio-seismic), neon and argon (coronal). 
}
\end{table*}

\begin{figure*}
\includegraphics[width=.75\textwidth]{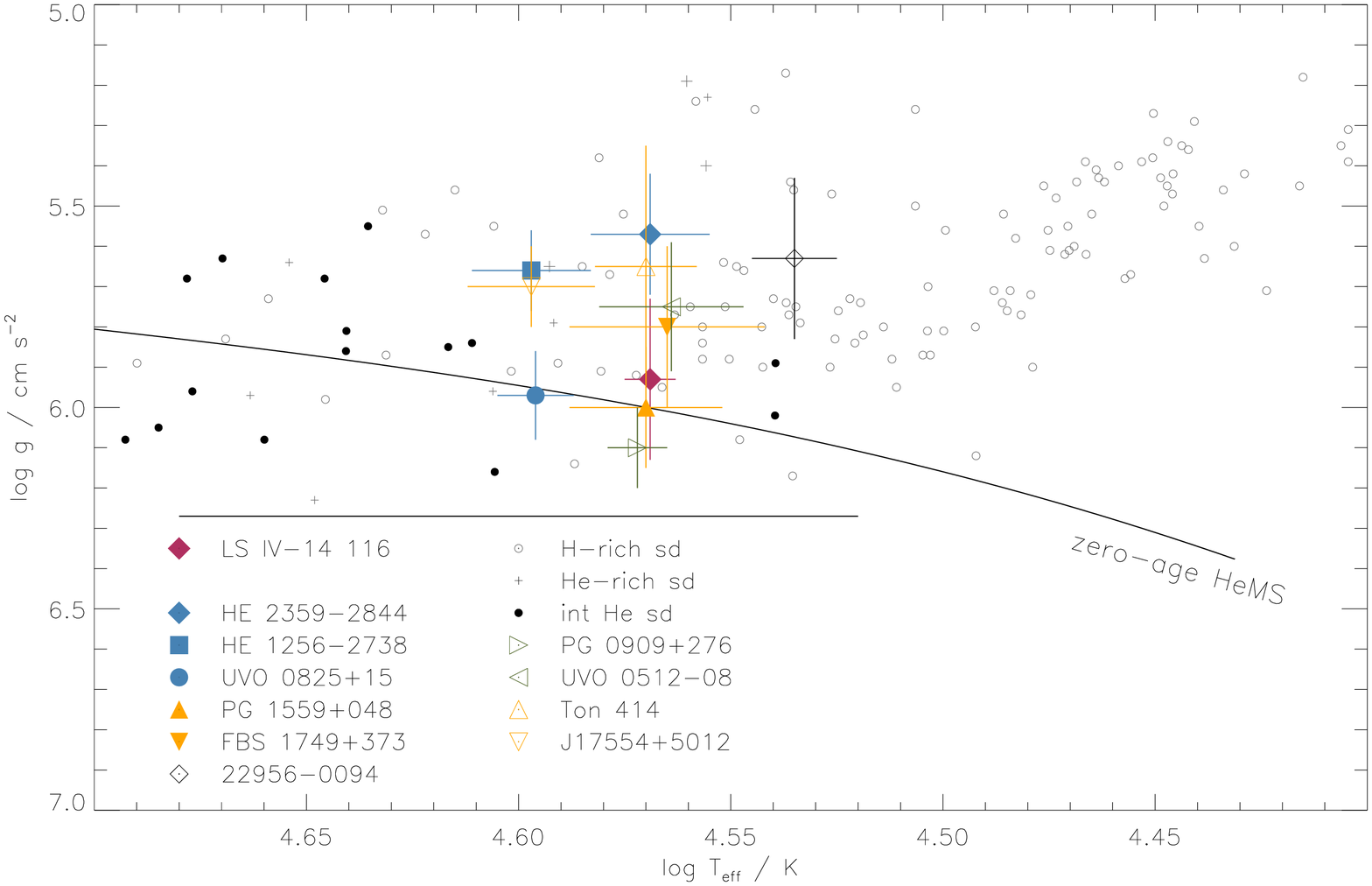}\\
\includegraphics[width=.75\textwidth]{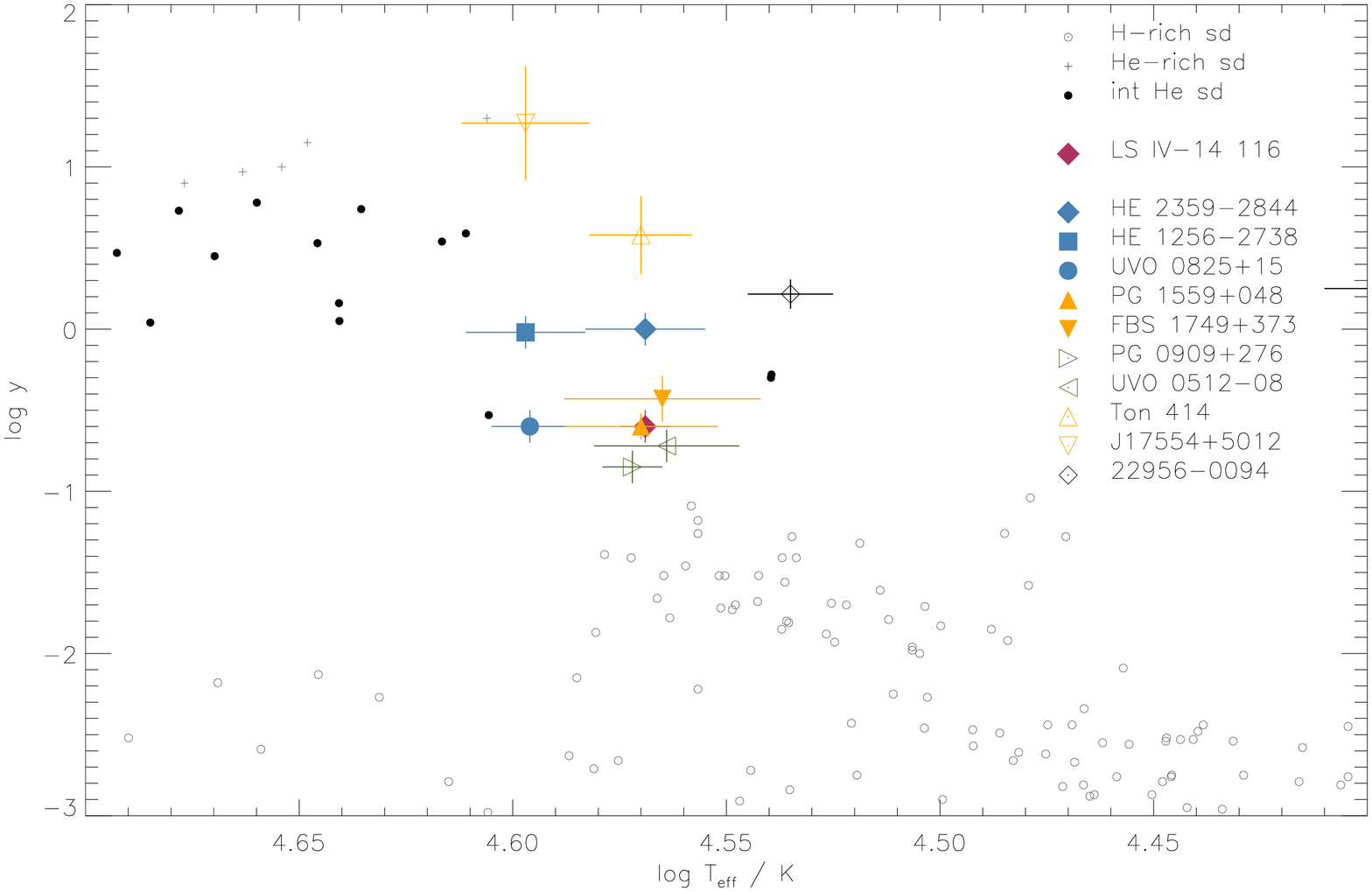}
\caption{The distribution of  heavy-metal (filled coloured symbols), helium-rich and normal hot subdwarfs with effective temperature, surface gravity (top) and surface helium-to-hydrogen ratio (bottom). In the $g-\teff$ plot, the solid line shows representative positions for the theoretical zero-age helium main-sequence (HeMS: $Z=0.02$). The observed data are from this paper (navy), \citet{naslim11,naslim13,jeffery17a,nemeth12}  and \protect\citet{wild17}. Filled symbols represent heavy-metal subdwarfs. 
}
\label{fig:teff_logg}
\end{figure*}

\section[]{Atmospheric parameters}
 
We measured the atmospheric  parameters effective temperature $T_{\rm eff}$, surface gravity $g$, and fractional helium abundance $n_{\rm He}$ of each star by fitting the observed spectrum with a grid of theoretical spectra. 
We used  the same $\chi^{2}$-minimization method as \citet{naslim10, naslim11, jeffery13, jeffery17a}. 
The  method compares the observed spectrum with theoretical emergent spectra computed from a grid of fully-blanketed plane-parallel model atmospheres in local thermodynamic and hydrostatic equilibrium \citep{behara06} using the Armagh optimisation code {\sc sfit} \citep{jeffery01b}. {\sc sfit} can use the entire observed spectrum but, by allowing the user to adjust the error associated with each datum in selected wavelength windows, can be tuned to exclude bad pixels, interstellar lines or other aggravating features. In fitting the entire blue-optical spectra of hot subdwarfs, the Stark-broadened helium and hydrogen line profiles dominate the measurement of $T_{\rm eff}$, $\log g$, and $n_{\rm He}$, although at high resolution there is a minor contribution to the $\chi^2$ surface from the ionization equilibria of subordinate species. 

The model atmosphere grid covers the range $T_{\rm eff} = 32\,000 (2000) 42\,000$ K, $\lgcs = 5.4 (0.2) 6.2$ and $n_{\rm He} =0.1, 0.2, 0.299, 0.699, 0.949$ and 0.989 where parentheses represent the grid step sizes. 
In practice, a subset of the full grid was used in order to meet computer memory limitations; the grid subset was always adjusted iteratively to ensure that its boundaries bracketed the final solution for a given star. 

Two choices were used for the distribution of elements heavier than helium in the models, namely i) {\bf m10}: 1/10 solar for all $Z>2$ ($\equiv [X/{\rm H}]=-1$) and ii) {\bf sdb}: 1/10 solar for $2<Z<26$ and solar for $Z\geq26$.  For two iHe-sd stars in our sample, PG\,1559+048 and FBS\,1749+373 which show trans-iron elements in their observed spectra, we determine the abundances and atmospheric parameters using both {\bf m10} and {\bf sdb} grids for comparison. 
The choice of grid ({\bf m10} or {\bf sdb}) made very little difference to the derived surface abundances  (see \S\,4). 
 
We determined $T_{\rm eff}$, $\log g$, and $n_{\rm He}$ by fitting the entire spectral region 3900--5000\,\AA\ using {\sc SFIT}. 
We excluded the region below $<3900$\AA\ where normalization is difficult due to blending between high-order diffuse lines in the Balmer and helium series.

Table \ref{t_pars} lists our measurements of $T_{\rm eff}$, $\log g$, and $n_{\rm He}$. 
The differences between measurements made using  grids ({\bf m10} and {\bf sdb}) are very small compared to the formal errors.
The projected equatorial rotation velocity (\vsini) was measured by optimizing fits to the profiles of carbon and nitrogen lines.   
Table \ref{t_pars} also includes previous measurements of \Teff, \sg\ and \nHe\ for all four stars.  
These were based on low-resolution optical spectra and non-LTE model atmospheres with line blanketing due to H and He only \citep{thejll94,nemeth12}.

Figure\,\ref{fig:teff_logg} shows the distribution of sample stars and of other heavy-metal and chemically-peculiar subdwarfs in the $g-\Teff$ and $y-\Teff$ planes.

 Formal errors in $T_{\rm eff}$ were obtained by fixing $\log g$ and $n_{\rm He}$ and fitting the spectral region 4680--4720\,\AA\ which includes the temperature sensitive He\,{\sc ii} 4686\,\AA\, and He\,{\sc i} 4713\,\AA\ lines. We found formal errors in $\log g$ by fitting H${\epsilon}$, H${\delta}$ and He {\sc i} 4026\,\AA\ separately while holding $T_{\rm eff}$ and $n_{\rm He}$ at fixed values.

\begin{figure*}
\includegraphics[height=.80\textheight]{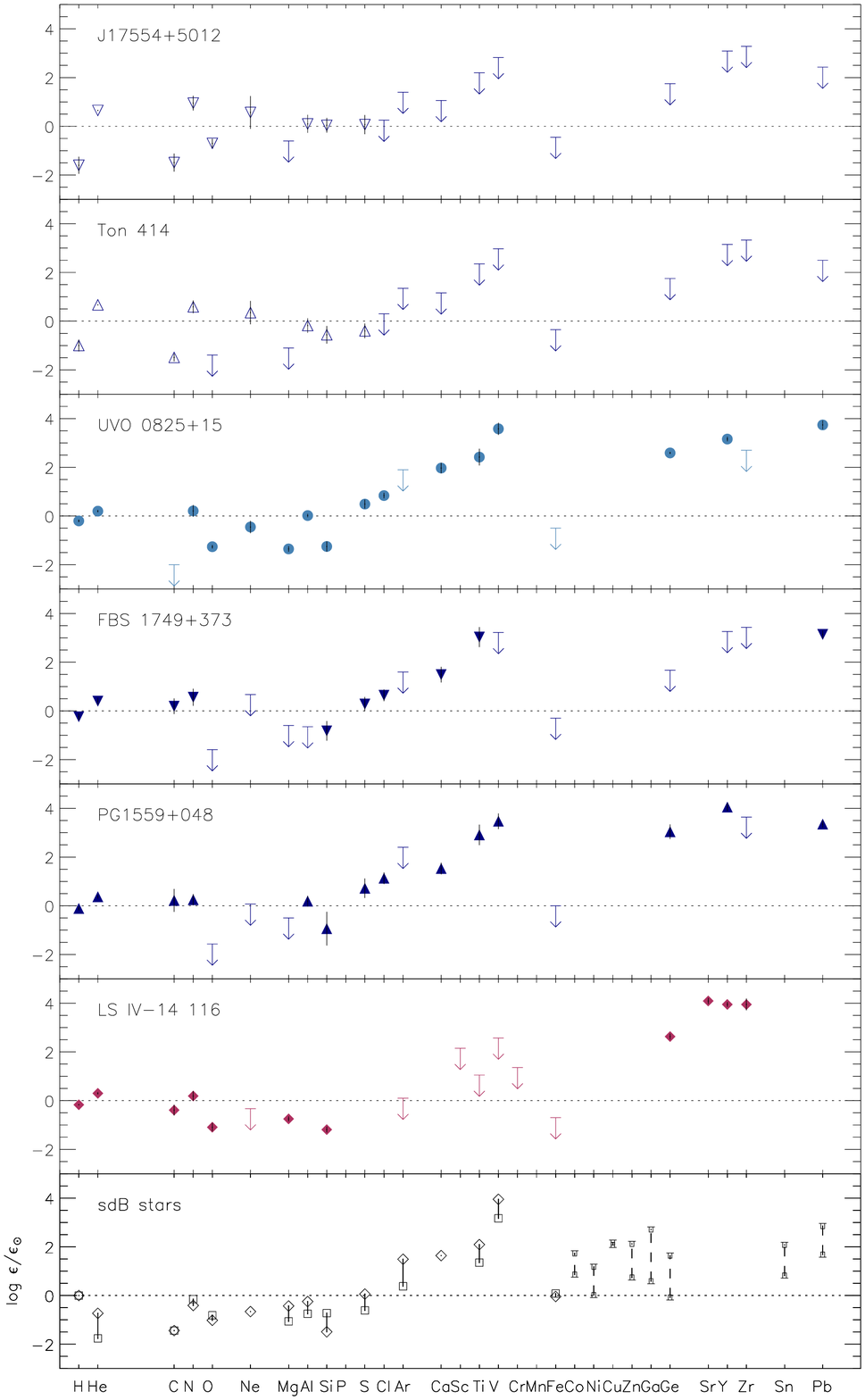}
\caption{Surface abundances of heavy-metal subdwarfs (filled symbols), helium-rich subdwarfs (open triangles) and normal subdwarfs (bottom panel). Abundances are shown relative to solar values (dotted line). Abundance ranges for normal subdwarf B stars are shown for  (i) $Z\leq26$ (solid lines): the average abundances for cool (squares) and warm (diamonds) sdBs \protect\citep{geier13} and (ii) $Z\geq27$ (broken lines): the range (delimited by hats) of abundances measured for five normal sdBs from UV spectroscopy \protect\citep{otoole06}. 
}
\label{fig:abs}
\end{figure*}

\begin{figure*}
\centering
\includegraphics[scale=0.55]{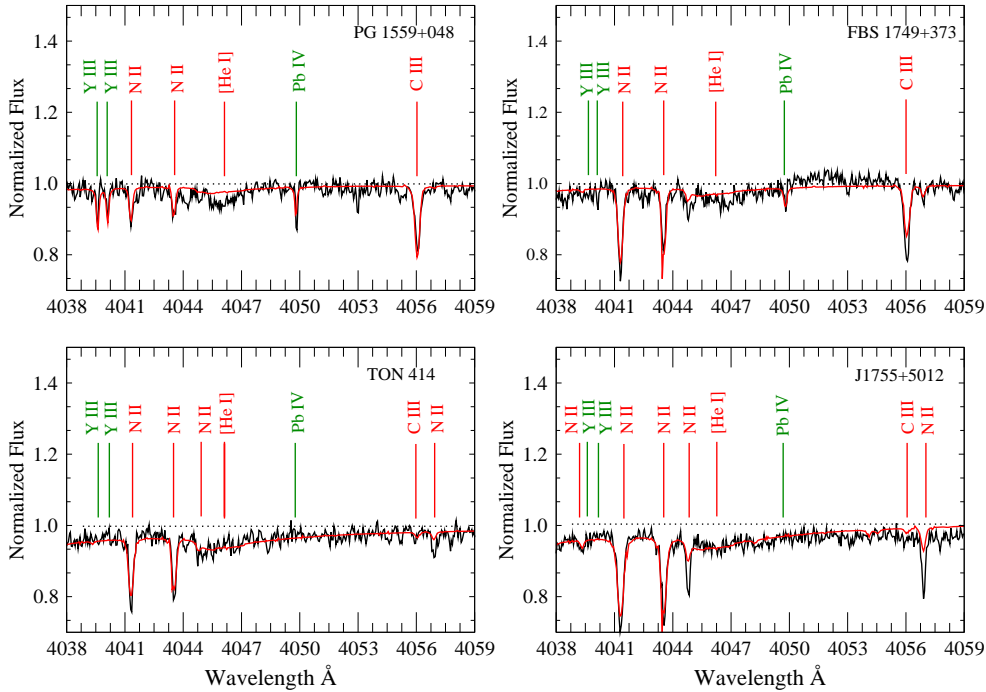}\\
\caption{Locations of Pb\,{\sc iv} 4049.48\,\AA,  Y\,{\sc iii} 4039.60\,\AA, and Y\,{\sc iii} 4040.11\,\AA\ in the Subaru spectra of PG\,1559+048, FBS\,1749+373, Ton\,414 and J17554+5012 along with the best fit model for each star. The Pb\,{\sc iv}  line is detected in both PG\,1559+048 and FBS\,1749+373. The Y\,{\sc iii} lines are detected only in PG\,1559+048. The broad depression around 4045.2\,\AA\ is due to a forbidden transition of He\,{\sc i} ($2^3 {\rm P}^0 - 5^3 {\rm P}^0$). Spectra are plotted in the {\it observer's} rest frame, thus include solar and stellar radial velocity shifts. }
\label{fitPb}
\end{figure*}

\section[]{Abundances}

We measured the equivalent widths of all C, N, O, Al, Si, S, Cl, Ca, Ti, V, Ge, Y and Pb absorption lines for which we had atomic data. 
Line abundances were calculated from equivalent widths using the LTE radiative transfer code {\sc spectrum} \citep{jeffery01b}. 
An atomic abundance is calculated from a given model atmosphere structure and a line equivalent width by Newton-Raphson iteration on the curve of growth.
Individual absorption line equivalent widths and abundances are shown in Table\,\ref{t_lines}. 

The mean surface abundances of measured elements in PG\,1559+048, FBS\,1749+373, Ton\,414 and J17554+5012 are shown in Table \ref{t:abs} in the form $\log \epsilon_i = \log n_i + c$ where $\log \Sigma_i a_i n_i = 12.15$ and $a_i$ are atomic weights.
This form conserves values of $\epsilon_i$ for elements whose abundances do not change, even when the mean atomic mass of the mixture changes substantially.

For PG\,1559+048 and FBS\,1749+373, which show lines due to trans-iron elements, we determined elemental abundances using models computed with both  mixtures {\bf m10} and {\bf sdb}. 
The differences between the abundances derived using these models are negligible (Table \ref{t:abs}). 
For the more helium-rich subdwarfs, Ton\,414 and J17554+5012, we determined abundances using mixture {\bf m10}.
In all cases, we used the grid model closest to the measured $T_{\rm eff}$, $\log g$, and $n_{\rm He}$ given in Table \ref{t_pars}. 
Full details of the adopted models are shown in the footnotes to Table\,\ref{t:abs} and Table\,\ref{t_lines}.

 In all cases we adopted a micro-turbulent velocity $v_{\rm t}=5\,{\rm km\,s^{-1}}$. 
Measuring $v{\rm_t}$ in iHe-sds is difficult as it a) requires an ion with (preferably) many lines covering a large range of equivalent width and b) generally assumes the photosphere to be chemically homogeneous in the parent atom. For the iHe-sds in our sample the condition is not satisfied and the photosphere is likely to chemically stratified by radiative levitation.  
 To test the consequences of adopting $v_{\rm t}=5\,{\rm km\,s^{-1}}$ we compared abundances for all four stars for $v_{\rm t}=0$ and $5\,{\rm km\,s^{-1}}$ and found  differences negligible compared with the statistical errors (Table\,\ref{t_vturb}). 
Reducing $v{\rm_t}$ to $0\,{\rm km\,s^{-1}}$ increases elemental abundances by between .01 and .09 dex.  

The errors given in {\bf Table \ref{t_lines}} are standard deviations of the line abundances about the mean or, in the case of a single representative line, derived from the estimated error in the equivalent width measurement. 

An additional systematic error is introduced by the errors in \Teff\ and $g$. 
\citet{naslim11} obtained representative values for $\delta\epsilon_i/\delta\Teff$,  $\delta\epsilon_i/\delta\log g$ and $\delta\epsilon_i/\delta v_{\rm t}$ for \crimson\ (ibid. Table 5). 
The latter are comparable with  values obtained for the current sample (Table\,\ref{t_vturb}). 
Note that \Teff, $\log g$ and $n_{\rm He}$ are correlated. 
For a given spectrum, similar fits can be obtained by increasing both \sg\ and \Teff\ simultaneously since higher gravity corresponds to higher pressure in the photosphere and hence to increased ionization. Similarly, raising $\teff$ requires more helium to match the He{\sc i} lines. Hence only $\delta\Teff$ need be considered.  
The abundance error due to the error in \Teff\ can then be computed from $\delta\Teff$ in Table\,\ref{t_pars} and  $\delta\epsilon_i/\delta\Teff$ from \citet[][Table 5]{naslim11}, supplemented for Ne, Al, Cl, Al, Ti, V, and Pb with error gradients measured from the current data.  This error should be combined quadratically with the error in Table\,\ref{t:abs} and a contribution due to the micro-turbulent velocity assumption from Table\,\ref{t_vturb}. 
Considering yttrium in PG\,1559+04, we find contributions of $\pm0.13$, $\pm0.10$ and $\pm0.01$ respectively giving a total error  $\delta\epsilon_{\rm Y}\approx \pm0.16$, whilst for carbon, the same calculation gives $\pm0.08$, $\pm0.46$ and $\pm0.08$ yielding $\delta\epsilon_{\rm C}\pm0.47$. 
Where the abundance is derived from few lines of a single ion with small scatter, the systematics are significant. In other cases, the statistical errors dominate. 

The errors given in Table \ref{t:abs} are the total errors derived from the quadratic sum of the statistical error and the systematic errors in \Teff\ and $v_{\rm t}$.

Surface abundances relative to the Sun for the programme stars, UVO\,0825$+$15 and LS\,IV$-14^{\circ}116$ are shown in Figure\,\ref{fig:abs} where they are compared with the observed range in surface abundance seen in normal subdwarf B stars \citep{geier13,otoole06}. 

 Figure\,\ref{fitPb} shows the Pb\,{\sc iv} 4049.48\,{\rm \AA} line in PG\,1559+048 and FBS\,1749+373. It also shows the  Y\,{\sc iii} 4039.60 and 4040.11\,{\rm \AA} lines in PG\,1559+048 together with the best fit model spectrum. Locations of these lines in Ton\,414 and J17554+5012 are shown for comparison. 
In PG\,1559+048 germanium is 3.0\,dex overabundant, lead is 3.3\,dex and yttrium is 4.0\,dex overabundant, relative to solar. In FBS\,1749+373, lead is 3.1\,dex overabundant relative to solar.
In both PG\,1559+048 and FBS\,1749+373 carbon and nitrogen are nearly solar or slightly over abundant, whereas oxygen, neon and aluminium are underabundant. 
 The He-sds, Ton\,414 and J17554+5012 do not show any elements heavier than sulphur. In both these stars carbon and oxygen are underabundant, while nitrogen is overabundant (0.6--0.9\,dex) relative to solar. Upper limits were estimated for significant elements not seen in the spectrum by assuming the equivalent width of the strongest line due to that element in the observed spectral range to be less than 5\,m\AA\,.

\begin{figure*}
\centering
\includegraphics[scale=0.45]{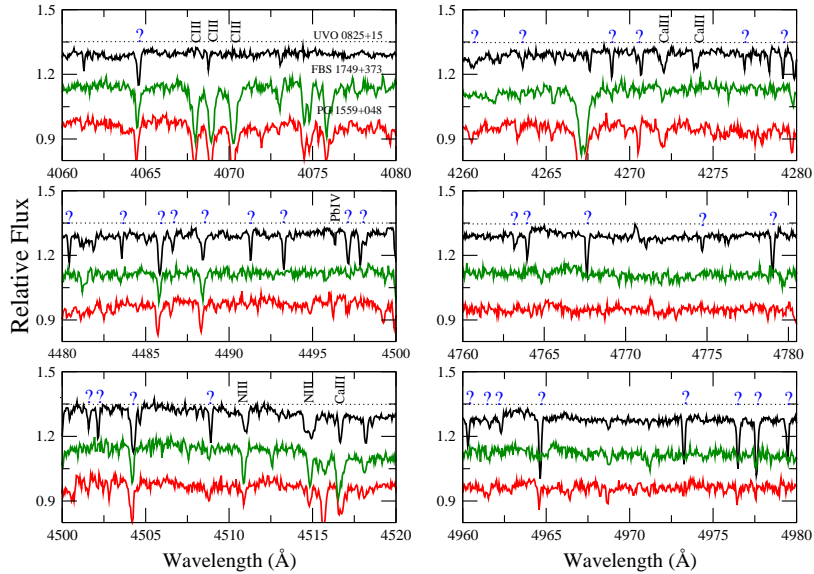}

\caption{Unidentified lines in Subaru HDS spectra of three iHe-sd stars UVO\,0825$+$15 \citep{jeffery17a}, FBS\,1749+373 and PG\,1559+048 (this paper).}
\label{lines_unid}
\end{figure*}

\citet{jeffery17a} reported over 150 unidentified absorption lines in the Subaru/HDS spectrum of UVO\,0825$+$15. The majority of these lines having equivalent widths above the detection threshold of 5\,m\AA\, are found in the Subaru/HDS spectra of lead-rich stars PG\,1559+048 and FBS\,1749+373 (Figure~\ref{lines_unid}). We cross checked these lines with the NIST Atomic Spectra Database; however these lines remain unidentified. 

\section{Conclusion}

Five He-sd stars were observed using the high-dispersion spectrograph on the Subaru telescope in an experiment aimed at discovering the numerical frequency of heavy-metal subdwarfs and the range in \Teff\, $g$ and $n_{\rm He}$ over which they occur.  
We have presented abundance analyses of two He-sd stars having helium-to-hydrogen ratio $\log y > 0.5$ and two iHe-sd stars having $-1 < \log y < -0.5$. The latter, PG\,1559+048 and FBS\,1749+373, show absorption lines due to triply ionized lead (Pb\,{\sc iv}) and have lead abundances between 3 and 4 dex above solar. PG\,1559+048 also shows germanium and yttrium (Ge\,{\sc iii} and Y\,{\sc iii}) with abundances comparable to those seen in LS\,IV$-14^{\circ}116$ \citep{naslim11}.
An analysis of the fifth star, and the third lead-rich star, was presented by \citet{jeffery17a}. Ton\,414 and J17554+5012 do not show absorption lines due to heavy elements germanium, yttrium or lead. Our analysis confirm that J17554+5012 is a 'helium enriched' He-sd having $\log y$= 1.27. With $\log y$= 0.58, Ton\,414 is strictly an iHe-sd, but has substantially more helium than the other two stars in which heavy elements were detected.

 From the sample of five stars, the three stars known to be iHe-sds (intermediate helium-rich subdwarfs) were discovered to be lead-rich. 
The other two were not. From this perspective it appears to be  possible to predict with a relatively high level of certainty which hot subdwarfs are likely to show trans-iron elements in their optical spectra. 
Testing this will be pursued with continuing observations elsewhere. 
From there, it will be a relatively straightforward step to evaluate the numerical frequency of heavy-metal subdwarfs using data on iHe-sds from current surveys such  as, for example, Gaia + LAMOST \citep{lei18,lei19}. 

  Our second goal was to explore the systematics of heavy-metal subdwarfs with regard to  distribution in  \Teff, $g$, and $y$. 
Figure\,\ref{fig:teff_logg} demonstrates that the heavy-metal subdwarfs analyzed so far cluster in the ranges $35\,000 < \Teff/{\rm K} < 40\,000$, $6.0 < \lgcs < 5.5$ and $-1 < \log y \simle 0$. Numbers remain too small to determine where zirconium, or lead, or both are most likely to be seen.  

 Corollaries to this question include whether heavy-metal superabundances occur outside this region, whether abundances of trans-iron elements are elevated  to a lesser extent in adjacent regions, and whether the regions includes stars with less exotic surface compositions.   
Addressing the first point, the two stars in the present sample which are not `heavy-metal' stars have $\log y\geq0.5$. 
With $\log y>1$, J17554+5012 at least is an extreme helium-rich hot subdwarf. Both are nitrogen-rich and carbon-poor. 
In addition, \citet{naslim10} analysed a sample of 5 He-sds and 1 iHe-sd with $30\,000 < \Teff/{\rm K} < 40\,000$ and $6.0 < \lgcs < 4.5$ without noting significant numbers of unidentified lines (ibid. Figs. A1 -- A6). The spectra of these stars were rechecked following the publication of \citet{naslim11} and \citet{naslim13}: no lines due to heavy elements were found.
The absence of heavy metal absorption lines in Ton\,414 and J17554+5012 and of optical Ge\,{\sc iii}, Y\,{\sc iii} and Zr\,{\sc iv} lines in any star analyzed before their identification in \crimson\ argues strongly that subdwarfs with $\log y \simge 0$ have not, so far, shown evidence of {\bf extreme} heavy-metal overabundances.
 Evidence of  heavy-metal enhancement by up to 2--3 dex would require high-resolution ultraviolet spectroscopy. 
A search for heavy-element enhancements is continuing via the SALT survey of chemically peculiar subdwarfs \citep{jeffery17b,jeffery17c,jeffery19b}. 

Turning to other questions raised in the introduction, it is clear that whatever physics produces extreme overabundances of lead and other heavy elements must also be associated with intermediate helium enrichment. 
Assuming the physics responsible is selective radiative levitation and that the surface layers are initially homogeneous, a superabundance by $d$ dex of element $Z$ in a region of mass $m$ must be associated with depletion in a region of mass $\geq 10^d m$.  
This represents a significant constraint for normal helium-poor sdB stars, where germanium, tin and lead, at least, are known to be elevated above solar values by 2--3 dex \citep{chayer06,otoole06}. 
This constraint becomes even more severe for the heavy-metal subdwarfs. 
Some relief can be obtained by assuming that the stellar atmosphere is heavily stratified and that the enriched layer is confined to the line forming region. 
Since this has direct consequences for the measurement of abundances in the photosphere, for the model atmospheres and the assumptions made in the analysis,  self-consistent model atmospheres which can account for the levitation of lead and other exotic species are urgently required to test this conjecture and to deduce the overall masses of enriched and depleted material.  

Broader questions relate to the origin and internal structure of the heavy-metal subdwarfs. 
Do they represent the high-temperature limit of the normal sdB stars in which radiative levitation has produced an extreme chemistry?
Or do they originate from some other channel? 
\citet{saio19} argue, for example, that \crimson\ represents the stripped core of a 3\Msolar\ helium star.
Answers to these questions will be addressed by looking at a larger ensemble of stars including other recent discoveries, and including distance and proper-motion measurements from Gaia.  

Meanwhile, discoveries of other heavy-metal subdwarfs are being announced \citep{jeffery19b,dorsch19,latour19b}.
Self-consistent model atmospheres with chemical diffusion and radiative levitation which can treat these heavy elements are urgently needed. 
Investigations of the masses, luminosities and space motions, and of the internal structure and evolutionary status of the heavy-metal subdwarfs are also in progress. 

\section*{Acknowledgments}

This paper is based on data collected at Subaru Telescope, which is operated by the National Astronomical Observatory of Japan.

This research has made use of the SIMBAD database, operated at CDS, Strasbourg, France. 

The Armagh Observatory and Planetarium is funded by direct grant form the Northern Ireland Dept for Communities.
NN acknowledges start-up fund from United Arab Emirates University (Fund number 31S378).
\bibliographystyle{mnras}
\bibliography{ehe}

\appendix
\renewcommand\thefigure{A.\arabic{figure}} 
\renewcommand\thetable{A.\arabic{table}} 

\section{Spectral analysis}
\begin{figure*}
\centering
\includegraphics[trim={0 12cm 0 0cm}, clip, scale=0.25]{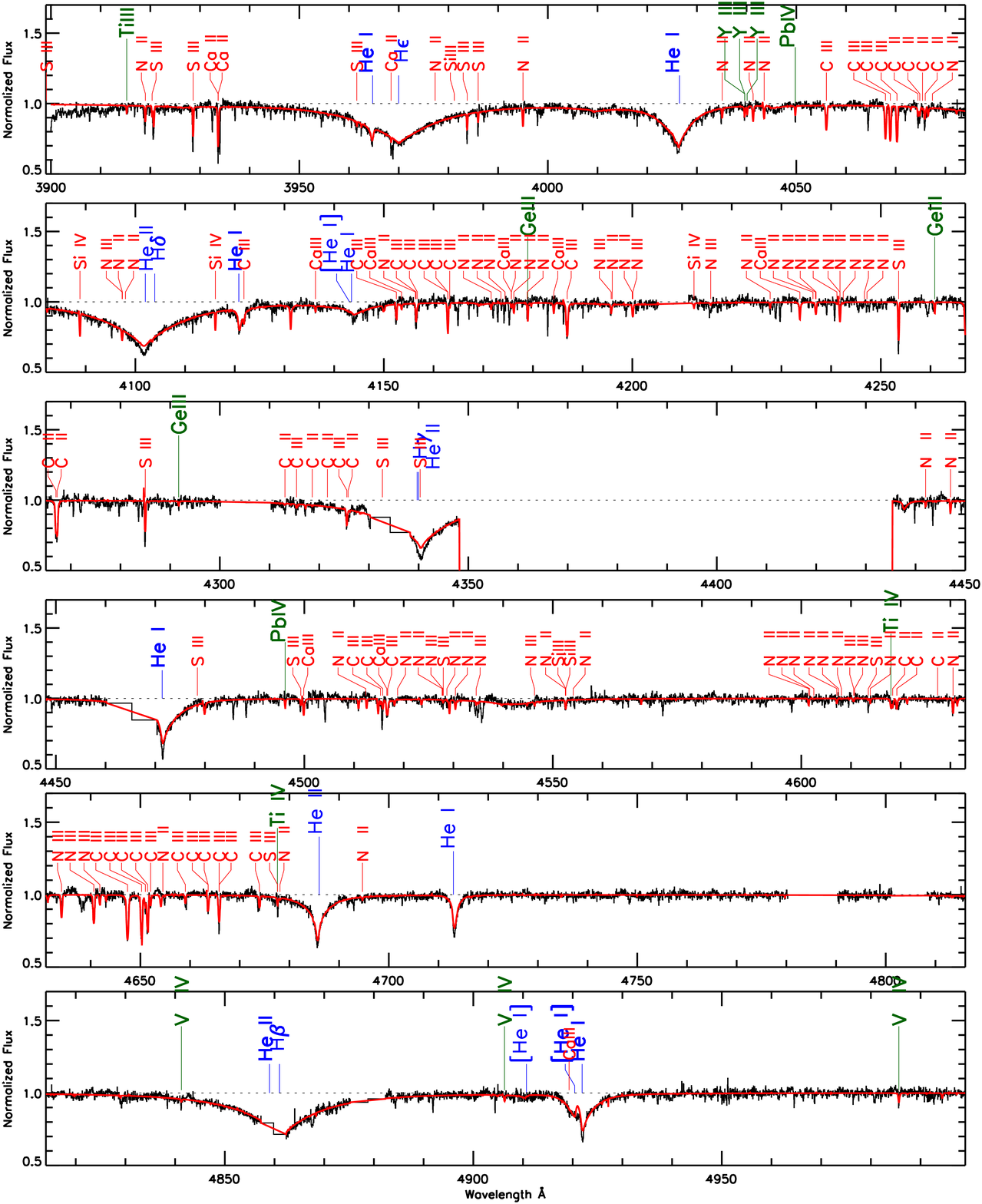}
\caption{ The Subaru spectrum of PG\,1559+048 along with the best fit model. }
\label{f:1559}
\end{figure*}

\begin{figure*}
\centering
\includegraphics[trim={0 12cm 0 0cm}, clip, scale=0.25]{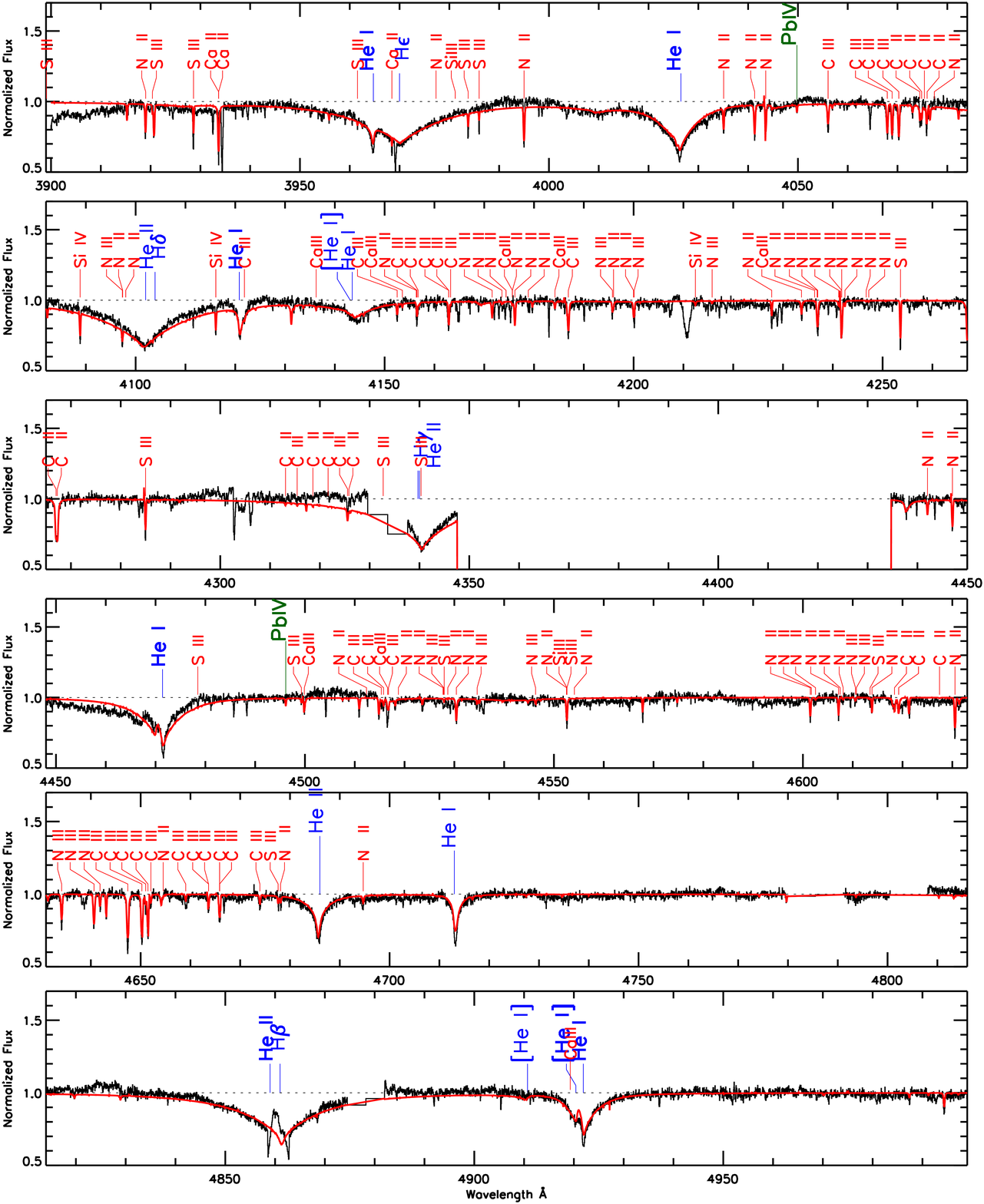}
\caption{The Subaru spectrum of FBS\,1749+373 along with the best fit model.}
\label{f:1751}
\end{figure*}

\begin{figure*}
\centering
\includegraphics[trim={0 12cm 0 0cm}, clip, scale=0.25]{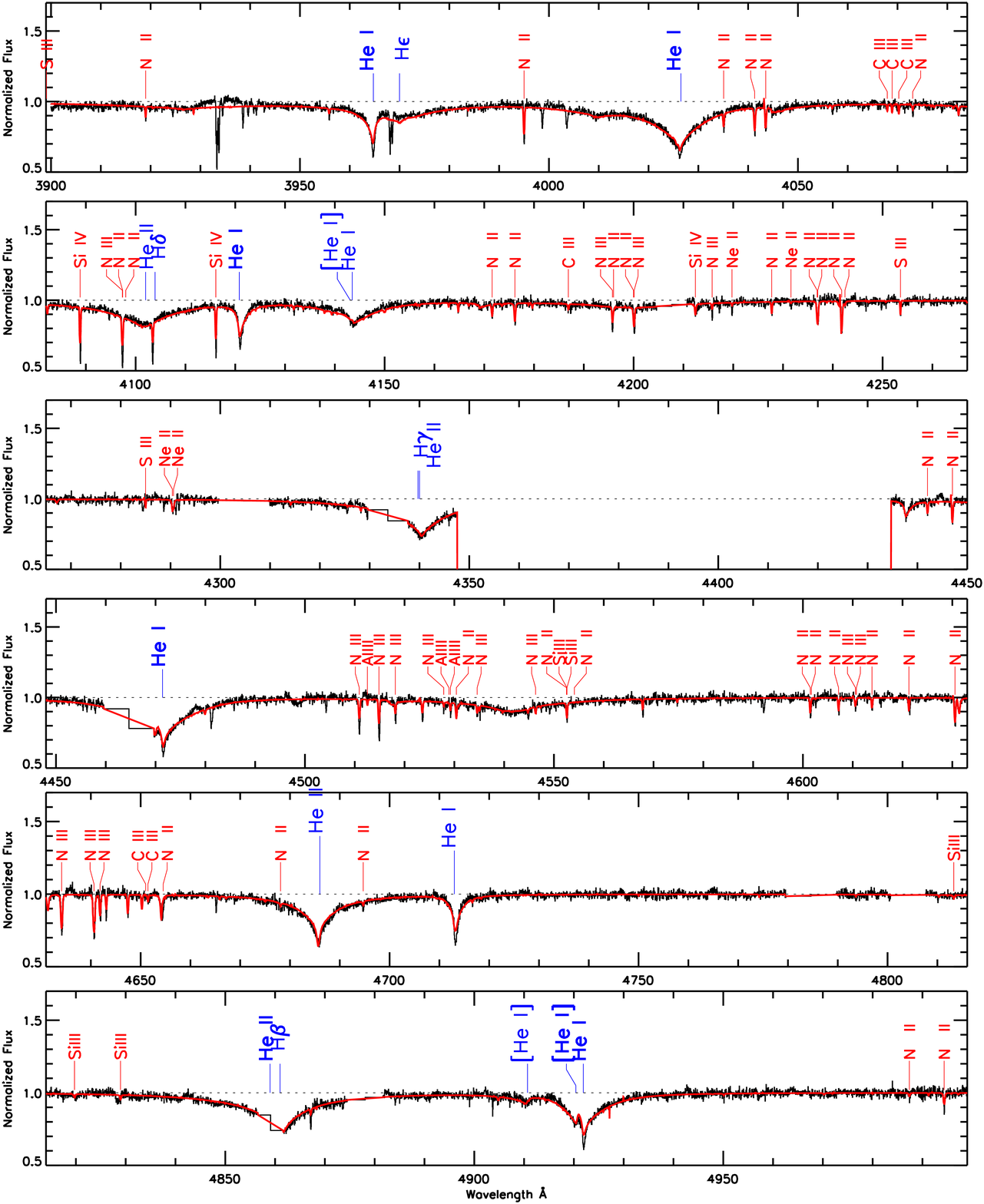}
\caption{The Subaru spectrum of Ton\,414 along with the best fit model.}
\label{f:0924}
\end{figure*}

\begin{figure*}
\centering
\includegraphics[trim={0 12cm 0 0cm}, clip, scale=0.25]{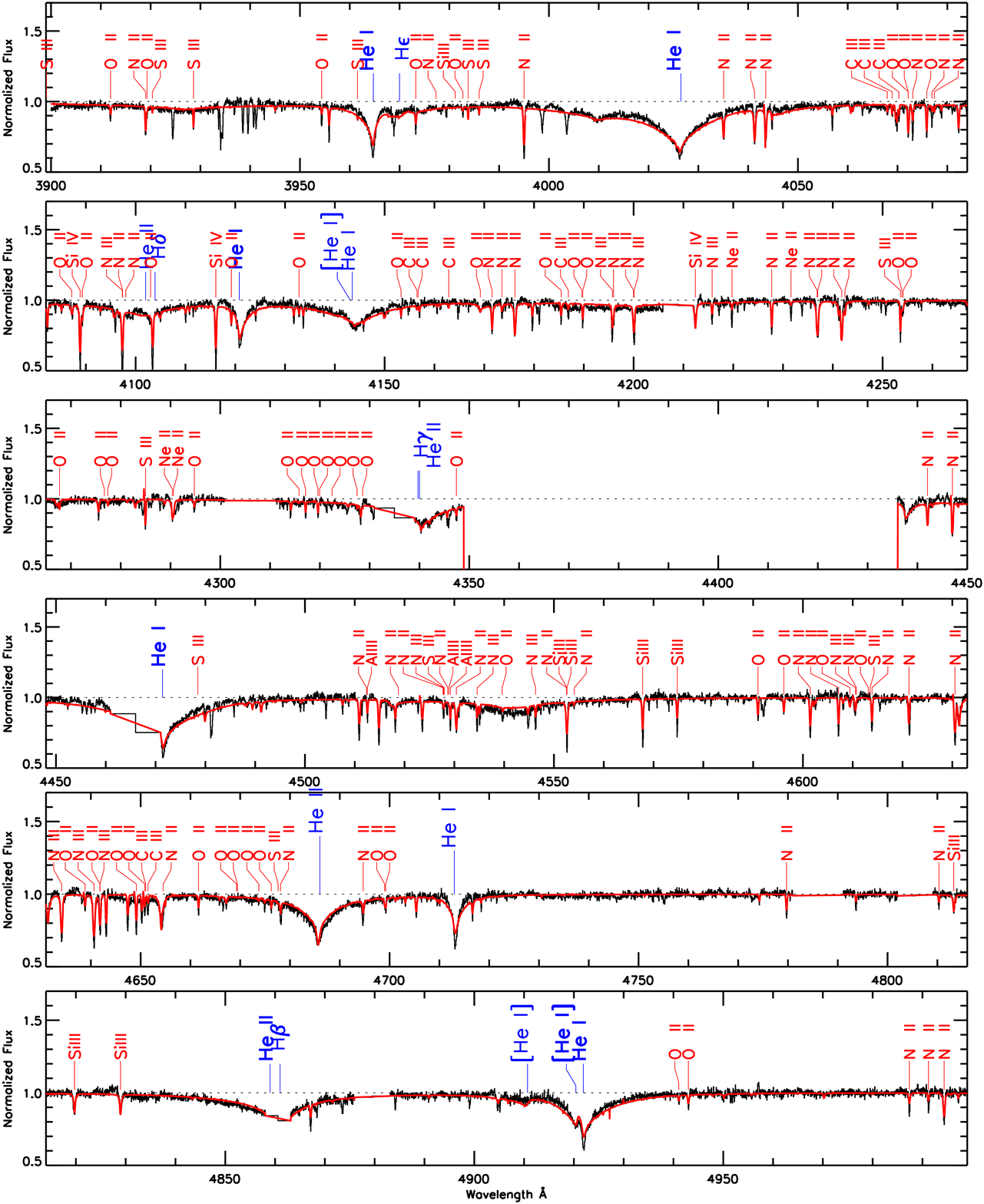}
\caption{The Subaru spectrum of J\,1755+5012 along with the best fit model.}
\label{f:1755}
\end{figure*}

This appendix contains the line-by-line analysis of the Subaru/HDS spectra, including equivalent widths and derived abundances (Table \ref{t_lines}) and a spectral atlas 
for each star (Figures~\ref{f:1559} -- \ref{f:1755}). Table \ref{t_vturb} compares abundances derived assuming a microturbulent velocity $v_t=0\kmsec$ with those presented in Table \ref{t:abs} assuming $v_t=5\kmsec$.


\begin{table*}
\setlength{\tabcolsep}{1pt} 
\centering
\caption{Measured equivalent width
  ($w_{\lambda}$) and derived abundances $\log \epsilon_i$ for each line. Averages and standard deviations (in parentheses) are reported for each ion. The model atmospheres and sources for $gf$-values used are reported at the end of the table.}
\label{t_lines}
\begin{tabular}{@{}ccccccccccccc}
\hline

\multicolumn{1}{c}{Ion} &
\multicolumn{3}{c}{PG\,1559+048} &
\multicolumn{3}{c}{FBS\,1749+373} &
\multicolumn{2}{c}{Ton\,414}&
\multicolumn{2}{c}{J\,1755+5012}\\
\multicolumn{1}{c}{$\lambda({\rm \AA})$} &

\multicolumn{1}{c}{$w_{\lambda}({\rm m\AA})$} &
\multicolumn{1}{c}{$^{a}\log \epsilon_{i}$} &
\multicolumn{1}{c}{$^{b}\log \epsilon_{i}$} &
\multicolumn{1}{c}{$w_{\lambda}({\rm m\AA})$} &
\multicolumn{1}{c}{$^{c}\log \epsilon_{i}$} &
\multicolumn{1}{c}{$^{d}\log \epsilon_{i}$} &
\multicolumn{1}{c}{$w_{\lambda}({\rm m\AA})$} &
\multicolumn{1}{c}{$^{e}\log \epsilon_{i}$} &
\multicolumn{1}{c}{$w_{\lambda}({\rm m\AA})$} &
\multicolumn{1}{c}{$^{f}\log \epsilon_{i}$}\\

\hline
  C\,{\sc ii}&      \multicolumn{9}{l}{}  \\  
 3918.98 &  60   &        9.19  & 9.13  &  53    &  8.68  & 8.68 &  &  &      &        \\ 
 3920.69 &  72   &        9.03  & 8.98  &  56    &  8.42  & 8.42  &   &   &       &        \\ 
 4074.48$\rceil$ & \\
 4074.52\,$\vert$&85   &  8.99  & 8.94  &  85    &  8.65  & 8.65  &   &   &       &         \\
 4074.85$\rfloor$&    \\
 4075.94$\rceil$ &   \\
 4075.85$\rfloor$&63   &  8.54  & 8.49  &  78    &  8.31  & 8.31  &   &   &       &        \\
 4267.02$\rceil$ &   \\
 4267.27$\rfloor$& 190   & 8.75  & 8.70  &  230   &  8.52  & 8.53  &   &   &       &        \\

         &       &  8.90(0.26) & 8.85(0.25)  &    &  8.52(0.16) & 8.52(0.16) &  &   &       &        \\[1mm]
C\,{\sc iii}     &       &      & &      &        & &     &         &        &        \\  
  4067.94  &  76   &  7.78  & 7.80  &  90    &  8.06  & 8.08 &    &         &   11     &   6.38     \\
  4068.91  &  85   &  7.87  & 7.89  &  98    &  8.14  & 8.16  &   &         &   30     &   6.88     \\
  4070.26  &  95   &  8.56  & 8.59  &  120    & 9.07  & 9.10  &   &         &  14      &  6.69      \\
  4647.42  &  160  &  8.46  & 8.47  &  195    & 8.73  & 8.75  &36  &  6.83  &   75     &  7.27      \\
  4650.25  &  120  &  8.31  & 8.34  &  146    & 8.60  & 8.63  &35  &  7.04  &   48     &  7.14      \\
  4651.47  &  102  &  8.60  & 8.62  &  128    & 8.93  & 8.95 &    &         &  30      &  7.30      \\
  4056.05  &  59   &  8.50  & 8.51  &  77    &  8.80  & 8.82  &   &         &        &        \\
  4152.51  &  100  &  9.14  & 9.17  &  29    &  8.72  & 8.74   &  &         &        &        \\
  4156.74 &  100  &  9.31  & 9.34  &  60    &  8.98  & 9.00   & &         &        &        \\
           &       &  8.50(0.50) & 8.52(0.51) &      &   8.67(0.35)&  8.69(0.36)   &      &   6.94(0.15)      &   &6.94(0.36)\\[2mm]
  N\,{\sc ii}         &       &   &    &      &        &  &    &         &        &       \\  
   3995.00  &   34      &  8.00 & 7.96 &78  &   8.30     & 8.31  &70   &    8.17     &   120     &  8.70      \\   
   4041.31  &   22      &   7.84& 7.80 &70  &   8.23     & 8.24  &70   &    8.13     &    97    &   8.30     \\ 
   4073.05  &           &       &      &    &            &       &33   &    8.65     &    57    &   8.90     \\
   4171.59  &           &       &      &    &            &       &40   &    8.34     &    60    &   8.51     \\
   4176.16  &           &       &      &    &            &       &42   &    8.05     &    97    &   8.52     \\
   4043.53  &   20      &  7.90 & 7.87 &50  &   8.13     & 8.13  &48   &    8.00     &   100     &  8.44      \\ 
   4236.98  &   22      &  8.14 & 8.10 &90  &   8.73     & 8.74  &80   &    8.53     &   140     &  8.91      \\  
   4241.78  &   36      &  8.24 & 8.21 &80  &   8.48     & 8.49  &67   &    8.25     &    90    &   8.39     \\   
   4447.03  &   30      &  8.25 & 8.22 &58  &   8.41     & 8.42  &62   &    8.43     &   100    &   8.86     \\ 
   4530.40  &           &       &      &    &            &       &58   &    8.27     &   105     &  8.62      \\ 
   4643.09  &           &       &      &    &            &       &50   &    8.59     &   97     &   9.16     \\ 
   4630.54  &   33    &  8.17   & 8.14 &72  &   8.44     & 8.46  &54   &    8.17     &  100      &  8.72      \\    
   4621.29  &         &         &      &    &            &       &    &              &    95    &  9.23      \\ 
   4613.87  &         &         &      &    &            &       &46   &    8.74     &    85    &   9.23     \\     
   4601.48  &   18    &  8.30   & 8.27 & 33 &   8.33     & 8.33  &50   &    8.58     &    92    &   9.09     \\ 
   4607.16  &         &         &      &    &            &   &37   &    8.46     &    83    &   9.08     \\ 
   4097.33  &         &        &       &51 &   7.57     &  7.59 &120   &   8.16      &  170      &  8.44      \\   

            &   &    8.10(0.17)&  8.07(0.17)  &        &    8.30(0.32)&  8.30(0.32)     &  &   8.35(0.23)    & & 8.77(0.31)    \\[1mm]
  N\,{\sc iii}       &       &  &     &  &           &       &     &    &        &        \\
    4640.64 &   77   &  8.09 & 8.14 &89   &   8.44    &  8.46 &  130    &   8.53      &    150    &  8.53\\
    4641.85 &        &       &      &    &            &       &   60   &    8.70     &    90    &   8.93     \\
    4103.43 &        &       &      &    &            &       &    90   &    8.14     &    150    &  8.60      \\
    4195.76 &        &       &      &    &            &       &   65   &    8.82     &     97   &9.01\\
    4200.10&    38   &  7.61 & 7.65 &65  &  8.92      &8.95 &    76    &     8.70      &     120    &   8.97     \\
    4634.14 &  30    &  8.29 & 8.32 &90    &  8.71      & 8.74 &  105  &    8.56     &     150   &  8.79      \\
    4610.55&         &       &   &   &               &         &     &                 &     50   &  9.07      \\
    4544.80 &         &       &   &   &               &        &30   &    8.60            &     42   &  8.63      \\
    4546.32&         &       &   &   &               &    &  &                &     42   &   8.65     \\
            &       &  7.99(0.35) &  8.03(0.35)  &      &       8.69(0.24)&   8.72(0.25)      &      &            8.58(0.22)    &    &  8.80(0.20)   \\[1mm]
                               
\hline
\end{tabular}
\end{table*}

\newpage
\addtocounter{table}{-1}
\begin{table*}
\setlength{\tabcolsep}{1pt} 
\centering
\caption{contd.}
\begin{tabular}{@{}ccccccccccccc}
\hline

\multicolumn{1}{c}{Ion} &
\multicolumn{3}{c}{PG\,1559+048} &
\multicolumn{3}{c}{FBS\,1749+373} &
\multicolumn{2}{c}{Ton\,414}&
\multicolumn{2}{c}{J\,1755+5012}\\
\multicolumn{1}{c}{$\lambda({\rm \AA})$} &

\multicolumn{1}{c}{$w_{\lambda}({\rm m\AA})$} &
\multicolumn{1}{c}{$^{a}\log \epsilon_{i}$} &
\multicolumn{1}{c}{$^{b}\log \epsilon_{i}$} &
\multicolumn{1}{c}{$w_{\lambda}({\rm m\AA})$} &
\multicolumn{1}{c}{$^{c}\log \epsilon_{i}$} &
\multicolumn{1}{c}{$^{d}\log \epsilon_{i}$} &
\multicolumn{1}{c}{$w_{\lambda}({\rm m\AA})$} &
\multicolumn{1}{c}{$^{e}\log \epsilon_{i}$} &
\multicolumn{1}{c}{$w_{\lambda}({\rm m\AA})$} &
\multicolumn{1}{c}{$^{f}\log \epsilon_{i}$}\\

\hline
  O\,{\sc ii}   &       &       &  &    &        &      &    &     &        &        \\  
   4649.14      &         &       & &     &        &      &   &       &    65    &  8.02      \\
   4072.15     &         &       &   &   &        &      &     &     &     44   &   7.82     \\
   4069.88     &         &       &   &   &        &      &      &    &     45   &  8.01      \\
   4075.86      &         &    & &   &      &        &      &         &   50    &  7.75      \\
   4119.21      &         &     & &  &      &        &      &         &   31   &  7.74      \\
   4189.78      &         &     & &  &      &        &      &         &   35   &  7.93      \\
   4590.97       &         &    & &   &      &        &      &        &   37   &  8.01      \\
   4596.17        &         &   & &    &      &        &      &        &  37    &   8.16     \\
   4638.85         &         &  & &     &      &        &      &        &  27    &  8.07      \\
   4650.84      &         &     & &  &      &        &      &        &   28   &  8.11      \\
   4661.63     &         &      & & &      &        &      &        &   35   &   8.17     \\
   4676.23      &         &     & &  &      &        &      &        &  25    &  8.08      \\
   4942.99       &         &    & &   &      &        &      &        &  28    & 8.06       \\
                &       &       & & &     &        &      &        &        &    7.99(0.14)     \\[1mm]
  Ne\,{\sc ii}  &       &       &  & &    &        &      &         &        &        \\  
  4219.76      &       &       &  & &    &        &   21   &   8.60      &   50     &   8.96     \\  
  4231.60      &       &       &  & &    &        &   16   &   8.77      &   41     &   9.16     \\  
  4290.37      &       &       &  & &    &        &   26   &   7.88      &   40     &   8.00     \\  
  4290.60      &       &       &  & &    &        &   26   &   7.88      &   30     &   7.86     \\ 
                     &       &    & &   &      &        &      &  8.28(0.47)      &        &   8.50(0.66)      \\[1mm]
  Al\,{\sc iii}       &       &       &  & &    &        &      &         &        &        \\  
   4512.54  &   20    &  6.72     & 6.70  &   &    &    &  11    &    6.08     &    43    &  6.78      \\  
   4529.20  &   24    &  6.56     & 6.54  &   &    &    &  35    &    6.45     &    32    &  6.32      \\ 
            &       &  6.64(0.11)     & 6.62(0.11)  &   &     &   &      &   6.27(0.26)     &        &    6.55(0.33)     \\[1mm]
  Si\,{\sc iii}    &       &       &  &    &     &   &      &         &        &        \\  
  4552.62&  38    &  7.17&  7.11 &59   &   7.02 & 7.03 &  34&  6.69  &   105     &   7.49     \\
   4567.82&       &       &      &    &         &      &25  &  6.74  &   116    &  7.81      \\
   4574.76&       &       &      &    &        &       &   &         &   65     &   7.79     \\
   4828.96&       &       &      &    &        &       &   &         &   85     &   7.57     \\
   4819.72&       &       &      &    &        &       &   &         &   85     &   7.67     \\
   4813.30&       &       &      &    &        &       &   &         &   50     &   7.48     \\
   3796.10&  27   &  7.10 & 7.04 &34  &   6.81 & 6.82 &23 &   6.58   &   83     &   7.46     \\
          &       &  7.14(0.05) & 7.08(0.05)   &    & 6.92(0.15) &   6.93(0.15)     &      &   6.67(0.08)      &          &   7.61(0.15)      \\[1mm]
  Si\,{\sc iv}       &       &    & &   &      &        &      &         &        &        \\ 
    4088.85 &   35  &  6.08 & 6.08 & 50   &  6.40 & 6.41  &   100  & 6.90     &   193   &   7.58     \\
    4116.10 &  17   &  5.94 & 5.94 & 40   &  6.52 & 6.54  &    84  &  7.00    &  175    &  7.77      \\
    4654.14 &       &       &      &      &       &       &  140   & 7.25      &  185    &   7.31     \\
    4212.41 &       &       &      &      &       &       &   95   &  7.52    &  65     &    7.14    \\
            &       &   6.01(0.10) &  6.01(0.10) &      &   6.46(0.08)  &  6.47(0.09) &      &     7.17(0.28)   &        &   7.45(0.28)      \\[1mm]
 S\,{\sc ii}       &       &       &      &      & &  &      &         &        &        \\ 
 4253.59   &  82     &  7.99 & 7.95   &  77    &   7.54 & 7.56   &   22   &   6.55      &   88     &   7.69     \\ 
 4284.99   &  62     &  7.96 & 7.93    & 60     &  7.58 &  7.59   &  20    &  6.80       &   70     &  7.72      \\ 
 4332.71   &  18     &  7.44 & 7.40   &  23    &   7.24 & 7.25   &      &         &        &        \\ 
 3717.78   &  61     &  7.81 & 7.77   &  50    &   7.32 & 7.32   &   10   &   6.48      &   25     &   6.89     \\ 
 3662.01   &  37     &  7.73 &  7.68  &  32    &   7.30 & 7.31   &   15   &   6.92      &   16     &   6.93     \\ 
             &       &  7.79(0.22) &  7.75(0.22)  &      &   7.40(0.15) & 7.40(0.15)   &      &     6.69(0.21)   &        &   7.31(0.45)      \\[1mm]
  S\,{\sc iii}       &       &       &  &&    &        &      &         &        &        \\ 
  3838.32  &   41    &  7.34 &  7.30  &   49   &   7.12 & 7.13   &  30    &  6.77       &   65     &   7.33     \\ 
  3837.80  &   32    &  7.86 & 7.82   &   28   &   7.43 & 7.44   &      &         &   20     &   7.19     \\ 
  3928.62  &  62     &  7.98 &  7.94  &   55   &   7.52 & 7.53   &      &         &   25     &   6.94     \\ 
  3983.77  &   42    &  7.96 & 7.92   &   46   &   7.68 & 7.68   &      &         &   22     &   7.16     \\ 
  3985.97  &   30    &  8.07 & 8.03   &   27   &   7.67 & 7.67   &      &         &   15     &   7.30     \\ 
  3631.99  &   61    &  7.92 & 7.89   &   50   &   7.43 & 7.44   &      &         &   20     &   6.68     \\ 
  3709.32  &   83    &  7.96 & 7.91   &   32   &   7.00 & 7.00   &  23    &   6.80&   55     &   7.23     \\ 
             &       &  7.87(0.24) & 7.83(0.24)   &      &   7.40(0.26)  & 7.41(0.26)  &      &     6.79(0.02)   &        &    7.12(0.23)     \\[1mm]
 
\hline
\end{tabular}
\end{table*}

\newpage
\addtocounter{table}{-1}
\begin{table*}
\centering
\caption{contd.}
\begin{tabular}{@{}cccccccccccc}
\hline

\multicolumn{1}{c}{Ion} &
\multicolumn{3}{c}{PG\,1559+048} &
\multicolumn{3}{c}{FBS\,1749+373} \\
\multicolumn{1}{c}{$\lambda({\rm \AA})$} &

\multicolumn{1}{c}{$w_{\lambda}({\rm m\AA})$} &
\multicolumn{1}{c}{$^{a}\log \epsilon_{i}$} &
\multicolumn{1}{c}{$^{b}\log \epsilon_{i}$} &
\multicolumn{1}{c}{$w_{\lambda}({\rm m\AA})$} &
\multicolumn{1}{c}{$^{c}\log \epsilon_{i}$} &
\multicolumn{1}{c}{$^{d}\log \epsilon_{i}$} \\

\hline
 Cl\,{\sc ii}       &       &   &      &        &      &   \\  
   3720.45 &  15     &   6.63 & 6.59  &   10   &   6.14  & 6.14   \\[1mm]
    &       &       &  & &    &        &          \\ 
 Ca\,{\sc iii}       &       &       &  & &    &        &          \\  
 3761.61&   15    &  7.80 & 7.79   &  20    &   7.91  & 7.91         \\  
 4153.57&   15    &  7.89 & 7.90   &  10    &   7.74  & 7.74      \\  
 4184.20&   24    &  7.87 & 7.88   &  16    &   7.70  & 7.71    \\  
4233.71&  40     &  7.79   & 7.80 &  29    &   7.67  &   7.68       \\  
4499.88&23       &  7.46   & 7.48 &  24    &   7.57  &   7.58       \\  
4516.59&  55     &  8.15   & 8.17 &  75    &   8.41  &   8.42        \\  
4919.28&10       &  7.96   & 7.97 &      &         &            \\  
5137.73&   10    &  8.05   & 8.07 &      &         &      \\  
       &         &   7.87(0.21)    & 7.88(0.20)   &      &  7.83(0.30)   &   7.84(0.30)                  \\[1mm]
Ti\,{\sc iii}       &       &     &  &      &    &           \\  
3915.26    &  18     &   8.00 & 7.95  &   20   &   7.70  &    7.70      \\  
Ti\,{\sc iv}       &       &     &  &      &    &           \\
4618.04    &  10     &   7.39 & 7.42  &      &       &       \\  
4677.59    &  25     &   8.19 & 8.22  &  20    &   8.25  &    8.27      \\      
           &         &   7.79(0.57)   & 7.82(0.56)    &        &    &           \\[1mm]
V\,{\sc iv}       &       &    &   &      &     &          \\  
4841.26    &  24     &  7.76 & 7.79   &      &     &          \\ 
4985.64   &   23    &   7.37 & 7.40  &      &      &          \\ 
5130.78   &   20    &  7.20  & 7.24  &      &      &         \\ 
5146.52   &   15    &  7.27  & 7.31  &      &      &          \\ 
          &         &  7.40(0.25) &  7.43(0.25)  &      &     &              \\[1mm]
Ge\,{\sc iii}       &       &   &    &      &      &         \\  
    4178.96         &  23     &  6.53 &  6.46  &      &    &           \\
    4291.71         &  10     &  6.84 &  6.77  &      &    &            \\
                    &         &  6.69(0.22)     &  6.62(0.22)  &      &    &              \\[1mm]
Y\,{\sc iii}       &       &       &  &    &        &        \\  
4039.60       &   15    &   6.02   & 6.00  &      &     \\ 
4040.11       &   20    &   6.16   & 6.13  &      &      \\ 
                &       &    6.09(0.10) & 6.07(0.09)  &   &    &             \\[1mm] 
Pb\,{\sc iv}       &       &       &      &  &     &         \\  
 3962.48           &  12   &  4.94   & 4.92 &  13    &   4.81     &  4.81    \\ 
 4049.48           &  20   &  5.16   & 5.14 &  19    &   4.97     &  4.97    \\ 
 4496.15           &  21   &  5.20   & 5.18 &      &         \\[1mm]
                   &       &    5.10(0.14)   & 5.08(0.14)  &      &      4.89(0.11) & 4.89(0.11)     \\[1mm]             
\hline
\end{tabular}
\parbox{170mm}{
$gf$ values were taken from the following. 
C\,{\sc ii}: \citet{yan87}, 
C\,{\sc iii}: \citep{hibbert76, hardorp70}, 
N\,{\sc ii}: \citet{becker90a}, 
N\,{\sc iii}: \citet{butler84},
O\,{\sc ii}: \citet{becker88b}, 
Ne\,{\sc ii}: \citet{wiese66}, 
Al\,{\sc iii}: \citet{cunto93}, 
Si\,{\sc iii}: \citet{becker90a}, 
Si\,{\sc iv}: \citet{becker90a}, 
S\,{\sc ii}: \citet{wiese69}, 
S\,{\sc iii}: \citet{wiese69}, 
Cl\,{\sc ii}: \citet{rod89}, 
Ca\,{\sc iii}: \citet{kurucz99}, 
Ti\,{\sc iii}: \citet{warner69}, 
V\,{\sc iv}: \citet{martin88}, 
Ge\,{\sc iii}: \citet{naslim11}, 
Y\,{\sc iii}: \citet{naslim11}, and 
Pb\,{\sc iv}: \citet{naslim13}
}\\[1mm]
\parbox{170mm}{
Model atmospheres:\\
$a$:   model: $T_{\rm eff}$ =  38\,000\,K, $\log$\,g=6.0, $n_{\rm He}$=0.200, {\bf m10}                    \\
$b$:   model: $T_{\rm eff}$ =  38\,000\,K, $\log$\,g=6.0, $n_{\rm He}$=0.200, {\bf sdb}                         \\
$c$:   model: $T_{\rm eff}$ =  36\,000\,K, $\log$\,g=6.0, $n_{\rm He}$=0.300, {\bf m10}                    \\
$d$:   model: $T_{\rm eff}$ =  36\,000\,K, $\log$\,g=6.0, $n_{\rm He}$=0.300, {\bf sdb}        \\
$e$:   model: $T_{\rm eff}$ =  38\,000\,K, $\log$\,g=6.0, $n_{\rm He}$=0.699, {\bf m10}                    \\
$f$:   model: $T_{\rm eff}$ =  40\,000\,K, $\log$\,g=6.0, $n_{\rm He}$=0.949, {\bf m10}                    \\
}
\end{table*}

\begin{table*}
\setlength{\tabcolsep}{6pt} 
\centering
\caption{Comparison of abundances $\log \epsilon_{i}$ for microturbulent velocities $v_t=0.0$ and 5.0\,km\,s$^{-1}$. Errors in parentheses are standard deviations about the mean.}
\label{t_vturb}
\begin{tabular}{@{}ccccccccc}
\hline
& \multicolumn{2}{c}{PG\,1559+048} &
\multicolumn{2}{c}{FBS\,1749+373} &
\multicolumn{2}{c}{Ton\,414} &
\multicolumn{2}{c}{J\,1755+5012}\\
$v_t/\kmsec$ & 0 & 5 &  0 & 5 &   0 & 5 &   0 & 5 \\   
Element &
$\log \epsilon_{i}$ &
$\log \epsilon_{i}$ &
$\log \epsilon_{i}$ &
$\log \epsilon_{i}$ &
$\log \epsilon_{i}$ &
$\log \epsilon_{i}$ &
$\log \epsilon_{i}$ &
$\log \epsilon_{i}$ \\
\hline
 C & 8.73(0.46) & 8.65(0.46) & 8.69(0.32) & 8.62(0.30) & 6.98(0.14) & 6.94(0.15) & 6.98(0.38) & 6.94(0.36) \\ 
 N & 8.12(0.22) & 8.08(0.22) & 8.47(0.35) & 8.39(0.34) & 8.48(0.25) & 8.42(0.25) & 8.86(0.30) &  8.78(0.28) \\
O  &            &            &            &            &            &            & 8.03(0.14) & 7.99(0.14)    \\  
Ne &            &            &            &            & 8.29(0.48) & 8.28(0.47) & 8.55(0.66) & 8.50(0.66)       \\  
Al & 6.69(0.11) & 6.64(0.11) &            &            & 6.30(0.28) & 6.27(0.26) & 6.60(0.33) & 6.55(0.33) \\
Si & 6.64(0.64) & 6.57(0.65) & 6.81(0.24) & 6.69(0.28) & 7.03(0.34) & 6.95(0.33) & 7.62(0.24) &  7.55(0.33) \\
 S & 7.98(0.28) & 7.84(0.23) & 7.53(0.27) & 7.40(0.21) & 6.77(0.17) & 6.72(0.17) &  7.28(0.42)& 7.19(0.32) \\
Cl & 6.62(0.06) & 6.63(0.06) & 6.15(0.20) & 6.14(0.20) & & & & \\
Ca & 7.88(0.21) & 7.87(0.21) & 7.84(0.31) & 7.83(0.30) & & & & \\
Ti & 7.87(0.42) & 7.86(0.42) & 7.99(0.38) & 7.98(0.38) & & & & \\
 V & 7.42(0.25) & 7.40(0.25) &     &      &    &  &  &         \\
Ge & 6.70(0.21) & 6.69(0.22) &     &      &    &  &  &          \\
 Y & 6.10(0.10) & 6.09(0.10) &     &      &    &  &  &          \\
Pb & 5.12(0.15) & 5.10(0.14) & 4.90(0.12) & 4.89(0.11) & & & & \\
\hline
\end{tabular}
\end{table*}

\label{lastpage}

\end{document}